\documentclass[12pt,a4paper,american,british,preprint]{iopart}
\usepackage[T1]{fontenc}
\usepackage[latin9]{inputenc}
\usepackage{color}
\usepackage{amsmath}
\usepackage{amssymb}
\usepackage{graphicx}
\usepackage{esint}
\PassOptionsToPackage{version=3}{mhchem}
\usepackage{mhchem}

\makeatletter

\pdfpageheight\paperheight
\pdfpagewidth\paperwidth

\usepackage{iopams}
\usepackage{setstack}

\expandafter\let\csname equation*\endcsname\relax 
\expandafter\let\csname endequation*\endcsname\relax 

\usepackage{babel}

\@namedef{ver@amsmath.sty}{}

\usepackage{amstext}

\makeatother

\usepackage{babel}
\begin{document}

\title{Optimized dynamical control of state transfer through noisy spin
chains}

\author{Analia Zwick, Gonzalo A. Álvarez, Guy Bensky and Gershon Kurizki}

\address{Weizmann Institute of Science, Rehovot 76100, Israel}
\begin{abstract}
We propose a method of optimally controlling the tradeoff of speed
and fidelity of state transfer through a noisy quantum channel (spin-chain).
This process is treated as qubit state-transfer through a fermionic
bath. We show that dynamical modulation of the boundary-qubits levels
can ensure state transfer with the best tradeoff of speed and fidelity.
This is achievable by dynamically optimizing the transmission spectrum
of the channel. The resulting optimal control is robust against both
static and fluctuating noise in the channel's spin-spin couplings.
It may also facilitate transfer in the presence of diagonal disorder
(on site energy noise) in the channel.
\end{abstract}
\maketitle
One dimensional (1D) chains of spin-$\frac{1}{2}$ systems with nearest-neighbor
couplings, nicknamed spin chains, constitute a paradigmatic quantum
many-body system of the Ising type \cite{ising_beitrag_1925}. As
such, spin chains are well suited for studying the transition from
quantum to classical transport and from mobility to localization of
excitations as a function of disorder and temperature \cite{kramer_localization:_1993}.
In the context of quantum information (QI), spin chains are envisioned
to form reliable quantum channels for QI transmission between nodes
(or blocks) \cite{bose_quantum_2003,Bose_review_2007}. Contenders
for the realization of high-fidelity QI transmission are spin chains
comprised of superconducting qubits \cite{lyakhov_quantum_2005,majer_coupling_2007},
cold atoms \cite{duan_controlling_2003,hartmann_effective_2007,fukuhara_quantum_2013,simon_quantum_2011},
nuclear spins in liquid- or solid-state NMR \cite{madi_time-resolved_1997,doronin_multiple-quantum_2000,zhang_simulation_2005,zhang_iterative_2007,cappellaro_dynamics_2007,rufeil-fiori_effective_2009,alvarez_perfect_2010,ajoy_algorithmic_2012},
quantum dots \cite{petrosyan_coherent_2006}, ion traps \cite{lanyon_universal_2011,blatt_quantum_2012}
and nitrogen-vacancy (NV) centers in diamond \cite{cappellaro_coherence_2009,neumann_quantum_2010,yao_scalable_2012,ping_practicality_2013}.

The distribution of coupling strengths between the spins that form
the quantum channel, determines the state transfer-fidelities \cite{bose_quantum_2003,zwick_quantum_2011,christandl_perfect_2005,karbach_spin_2005,kay_perfect_2006,kay_review_2010}.
Perfect state-transfer (PST) channels can be obtained by precisely
engineering each of those couplings \cite{christandl_perfect_2004,Albanese_mirror_2004,christandl_perfect_2005,karbach_spin_2005,kay_perfect_2006,paternostro_perfect_2008,kay_review_2010,zwick_robustness_2011}.
Such engineering is however highly challenging at present, being an
unfeasible task for long channels that possess a large number of control
parameters and are increasingly sensitive to imperfections as the
number of spins grows \cite{Alvarez_NMR_2010,zwick_robustness_2011,zwick_spin_2012,Zwick_Chapt_2013}.
A much simpler control may involve \textit{only} the boundary (source
and target) qubits that are connected via the channel. Recently, it
has been shown that if the boundary qubits are weakly-coupled to a
uniform (homogeneous) channel (\textit{i.e.}, one with identical couplings),
quantum states can be transmitted with arbitrarily high fidelity at
the expense of increasing the transfer time \cite{wojcik_unmodulated_2005,wojcik_multiuser_2007,Venuti_Qubit_2007,Venuti_Long-distance_2007,Giampolo_entanglement_2009,Giampaolo_Long-distance_2010,yao_robust_2011,zwick_spin_2012}.
Yet such slowdown of the transfer may be detrimental because of omnipresent
decoherence.

To overcome this problem, we here propose a hitherto unexplored approach
for optimizing the tradeoff between fidelity and speed of state-transfer
in quantum channels. This approach employs temporal modulation of
the couplings between the boundary qubits and the rest of the channel.
This kind of control has been considered before for a different purpose,
namely to implement an effective optimal encoding of the state to
be transferred \cite{Haselgrove_Optimal_2005}. Instead, we treat
this modulation as dynamical control of the boundary system which
is coupled to a fermionic bath that is treated as a source of noise.
The goal of our modulation is to realize an optimal spectral filter
\cite{clausen_bath-optimized_2010,clausen_task-optimized_2012,escher_optimized_2011,bensky_optimizing_2012,petrosyan_reversible_2009,gordon_universal_2007,gordon_optimal_2008,kofman_universal_2001,kofman_unified_2004}
that blocks transfer via those channel eigenmodes that are responsible
for noise-induced leakage of the QI \cite{wu_master_2009}. We show
that under optimal modulation, the fidelity and the speed of transfer
can be improved \textit{by several orders of magnitude}, and the fastest
possible transfer is achievable (for a given fidelity).

Our approach allows to reduce the complexity of a large system to
that of a simple and small open system where it is possible to apply
well developed tools of quantum control to optimize state transfer
with few universal control requirements on the source and target qubits.
In this picture, the complexity of the channel is simply embodied
by correlation functions in such a way that we obtain a universal,
simple, analytical expression for the optimal modulation. While in
this article we optimize the tradeoff between speed and fidelity so
as to avoid decoherence as much as possible, this description \cite{clausen_bath-optimized_2010,clausen_task-optimized_2012,escher_optimized_2011,bensky_optimizing_2012,petrosyan_reversible_2009,gordon_universal_2007,gordon_optimal_2008,kofman_universal_2001,kofman_unified_2004,wu_master_2009}
allows one to actively suppress decoherence and dissipation in a simple
manner, since it may be viewed as a generalization of dynamical decoupling
protocols \cite{Viola_Dynamical_1998,viola_dynamical_1999,Viola_RobustDD_2003,Lidar_QDynDec_2005}.
In what follows, we explicitly deal with a spin-chain quantum channel,
but point out that our control may be applicable to a broad variety
of other quantum channels.

\section{\label{sec:Quantum channel and state transfer fidelity}Quantum channel
and state transfer fidelity}

\subsection{Hamiltonian and boundary control}

We consider a chain of $N\!+\!2$ spin-$\frac{1}{2}$ particles with
XX interactions between nearest neighbors, which is a candidate for
a variety of state-transfer protocols \cite{bose_quantum_2003,Bose_review_2007,lyakhov_quantum_2005,majer_coupling_2007,duan_controlling_2003,hartmann_effective_2007,fukuhara_quantum_2013,simon_quantum_2011,madi_time-resolved_1997,zhang_simulation_2005,zhang_iterative_2007,cappellaro_dynamics_2007,rufeil-fiori_effective_2009,doronin_multiple-quantum_2000,ajoy_algorithmic_2012,alvarez_perfect_2010,petrosyan_coherent_2006,lanyon_universal_2011,blatt_quantum_2012,cappellaro_coherence_2009,neumann_quantum_2010,yao_scalable_2012,ping_practicality_2013,zwick_quantum_2011,christandl_perfect_2005,karbach_spin_2005,kay_perfect_2006,kay_review_2010,zwick_robustness_2011,christandl_perfect_2004,Albanese_mirror_2004,paternostro_perfect_2008}.
The Hamiltonian is given by
\begin{equation}
H=H_{0}+H_{bc}(t),\label{eq:hamiltonian}
\end{equation}
\begin{equation}
H_{0}=\sum_{i=1}^{N-1}\frac{J_{i}}{2}\left(\sigma_{i}^{x}\sigma_{i+1}^{x}+\sigma_{i}^{y}\sigma_{i+1}^{y}\right),\: H_{bc}(t)=\alpha(t)\sum_{i\in\{0,N\}}\frac{J_{i}}{2}\left(\sigma_{i}^{x}\sigma_{i+1}^{x}+\sigma_{i}^{y}\sigma_{i+1}^{y}\right),
\end{equation}
where $H_{0}$ and $H_{bc}$ stand for the chain and boundary-coupling
Hamiltonians, respectively, $\sigma_{i}^{x(y)}$ are the appropriate
Pauli matrices and $J_{i}$ are the corresponding exchange-interaction
couplings.

\subsection{Mapping to a few-body open-quantum system}

The magnetization-conserving Hamiltonian $H$ can be mapped onto a
non-interacting fermionic Hamiltonian \cite{lieb_two_1961} that has
the particle-conserving form 
\begin{equation}
H_{0}=\sum_{i=1}^{N-1}\frac{J_{i}}{2}\left(c_{i}^{\dagger}c_{i+1}+c_{i}c_{i+1}^{\dagger}\right),\: H_{bc}(t)=\alpha(t)\sum_{i\in\{0,N\}}\frac{J_{i}}{2}\left(c_{i}^{\dagger}c_{i+1}+c_{i}c_{i+1}^{\dagger}\right),
\end{equation}
where $c_{j}=\frac{1}{2}e^{i\frac{\pi}{4}\sum_{0}^{j-1}\sigma_{i}^{+}\sigma_{i}^{-}}\sigma_{j}^{-}$
create a fermion at site $j$ and $\sigma^{\pm}=\sigma^{x}\pm i\sigma^{y}$.
The Hamiltonian $H_{0}$ can be diagonalized as $H_{0}=\sum_{k=1}^{N}\omega_{k}b_{k}^{\dagger}b_{k}$,
where $b_{k}^{\dagger}=\sum_{j=1}^{N}\langle j|\omega_{k}\rangle c_{j}^{\dagger}$
populates a single-particle fermionic eigenstate $\vert\omega_{k}\rangle$
of energy $\omega_{k}$, and $\vert j\rangle=\vert0..01_{j}0..0\rangle$
denote the single-excitation subspace. Under the assumption of mirror
symmetry of the couplings with respect to the source and target qubits
$J_{i}=J_{N-i}$, the energies $\omega_{k}$ are not degenerate, $\omega_{k}<\omega_{k+1}$,
and the eigenvectors have a definite parity that alternates as $\omega_{k}$
increases \cite{karbach_spin_2005}. This property implies that $\langle j|\omega_{k}\rangle=(-1)^{k-1}\langle N-j+1|\omega_{k}\rangle$
and allows us to rewrite the boundary-coupling Hamiltonian as 
\begin{equation}
H_{bc}(t)=\alpha(t)J_{0}c_{0}^{\dagger}\underset{k=1}{\overset{N}{\sum}}\langle1|\omega_{k}\rangle b_{k}+\alpha(t)J_{N}c_{N+1}^{\dagger}\underset{k=1}{\overset{N}{\sum}}(-1)^{k-1}\langle N|\omega_{k}\rangle b_{k}+\mathrm{h.c.}
\end{equation}

For an odd $N$, there exists a single non-degenerate, zero-energy
fermionic mode in the quantum channel, labelled by $k=z=\frac{N+1}{2}$
\cite{wojcik_multiuser_2007,yao_robust_2011,ping_practicality_2013}.
As a consequence, the two boundary qubits ($0$ and $N+1$) are resonantly
coupled to this mode. Therefore, we consider these three resonant
fermionic modes as the ``system'' $S$ and reinterpret the other
fermionic modes as a ``bath'' $B$. In this picture, the system-bath
$SB$ interaction is off-resonant. Then, we rewrite the total Hamiltonian
as 
\begin{equation}
H=H_{S}(t)+H_{B}+H{}_{SB}(t),\label{eq:H}
\end{equation}
where 
\begin{equation}
H_{B}=\sum_{k\ne z,k=1}^{N}\omega_{k}b_{k}^{\dagger}b_{k},\: H_{S}(t)=s_{+}(t)\tilde{J}_{z}b_{z}+\mathrm{h.c.},
\end{equation}
\begin{equation}
H_{SB}(t)=s_{+}(t)\sum_{k\in k_{odd}}\tilde{J}_{k}b_{k}+s_{-}(t)\sum_{k\in k_{even}}\tilde{J}_{k}b_{k}+h.c.,\label{eq:Hsb}
\end{equation}
with $s_{\pm}(t)=\alpha(t)(c_{0}^{\dagger}\pm c_{N+1}^{\dagger})$,
$\tilde{J}_{k}=J_{1}\langle1\vert\omega_{k}\rangle$, $k_{odd}=\{1,3,..,N\},$
provided $k_{odd}\ne z$, and $k_{even}=\{2,4,..,N-1\}$.

The form (\ref{eq:H}) is amenable to the application of optimal dynamical
control of the multipartite system \cite{clausen_bath-optimized_2010,clausen_task-optimized_2012,gordon_scalability_2011,gordon_dynamical_2009,kurizki_universal_2013,Schulte-Herbruggen_Optimal_2009}:
such control would be a generalization of the single-qubit dynamical
control by modulation of the qubit levels \cite{escher_optimized_2011,bensky_optimizing_2012,petrosyan_reversible_2009,gordon_universal_2007,gordon_optimal_2008,kofman_universal_2001,kofman_unified_2004}.
To this end, we rewrite Eq. (\ref{eq:Hsb}) in the interaction picture
as a sum of tensor products between system $S_{j}$ and bath $B_{j}$
operators (see \ref{sec:Appendix-A:-Interaction}) 
\begin{equation}
H_{SB}^{I}(t)=\sum_{j=1}^{4}S_{j}(t)\otimes B_{j}^{^{\dagger}}(t).\label{eq:HSB_int-pict}
\end{equation}
From this form one can derive the system density matrix of the system,
$\rho_{S}(t)$, in the interaction picture, under the assumption of
weak system-bath interaction, to second order in $H_{SB}$, as \cite{clausen_bath-optimized_2010,escher_optimized_2011}
\begin{equation}
\rho_{S}(t)=\rho_{S}(0)-t\sum_{i,i'=1}^{6}R_{i,i'}(t)[\hat{\nu}_{i},\hat{\nu}_{i'}\rho_{S}(0)]+h.c.,\label{eq:rho_s}
\end{equation}
where 
\begin{equation}
R_{i,i'}(t)=\frac{1}{t}\sum_{j,j'=1}^{4}\int_{0}^{t}dt'\int_{0}^{t'}dt"\Phi_{j,j'}(t'-t")\Omega_{j,i}(t')\Omega_{j',i'}^{*}(t"),
\end{equation}
with $\Phi_{j,j^{'}}(\tau)=\mathrm{Tr}_{B}\left\{ B_{j}(\tau)B_{j^{'}}(0)\rho_{B}(0)\right\} $
denoting the correlation functions of bath operators and $\Omega_{j,i}(t)$
being a rotation-matrix in a chosen basis of operators $\hat{\nu}_{i}$
used to represent the evolving system operators, $S_{j}(t)=\underset{i=1}{\overset{6}{\sum}}\Omega_{j,i}(t)\hat{\nu}_{i}$
(\ref{sec:Appendix-A:-Interaction}). The solution (\ref{eq:rho_s})
will be used to calculate and optimize the state-transfer fidelity
in what follows.

\begin{figure}
\centering{}\includegraphics[width=0.7\columnwidth]{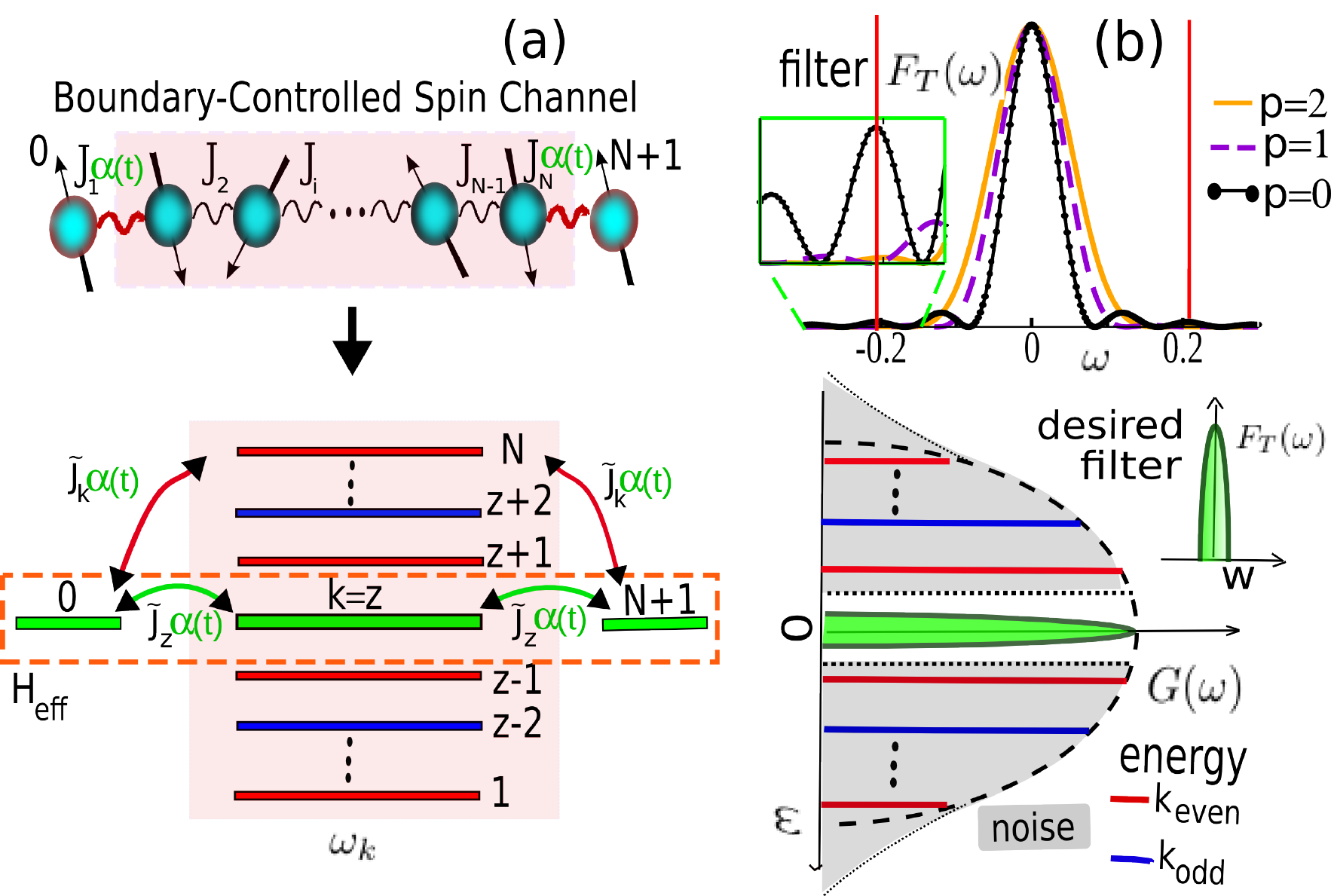}\caption{\label{fig:chain-FilterFunction}(Color online) (a) Top: State transfer
through a spin-channel with boundary-controlled couplings. Bottom:
Boundary-controlled spin chain mapped to a non-interacting spinless
fermionic system (dashed rectangle) which couples to the bath fermionic-modes
$k$ (red even $k$ and blue odd $k$ lines) with strengths $\tilde{J}_{k}\alpha(t)$.
The two boundary spins 0 and $N+1$ are resonantly coupled to the
chain by the fermionic-mode $z$ with a coupling strength $\tilde{J}_{z}\alpha(t)$
(green lines). (b) Bottom: Interacting bath-spectrum $G_{\pm}(\omega)$
including noise effects (grey color) bounded by the Wigner-semicircle
(maximal-disorder) lineshape (dashed contour) with a central gap around
$\omega_{z}$. In the central gap, an optimal spectral-filter $F_{T,\pm}(\omega)$
is shown (green color). Top: $F_{T,-}(\omega)$ generated by boundary-control
$\alpha_{p}(t)$: $p=0\mbox{ (black dotted), \ensuremath{p=1}(violet dashed), }p=2$
(orange thin). The red vertical lines are the nearest bath-spectrum
eigenenergies. The inset is a zoom on the tails of the filter spectrum
that protects the state transfer against a general noisy bath with
a central gap.}
\end{figure}

\subsection{Fidelity derivation}

We are interested in transferring a qubit state $\vert\psi_{0}\rangle$
initially stored on the $0$ qubit to the $N+1$ qubit . Here $\vert\psi_{0}\rangle$
is an arbitrary normalized superposition of the spin-down $\vert0_{0}\rangle$
and spin-up $\vert1_{0}\rangle$ (single-spin) states. To assess the
state transfer over time $T$, we calculate the averaged fidelity
$F(T)=\frac{f_{0,N+1}^{2}(T)}{6}+\frac{f_{0,N+1}(T)}{3}+\frac{1}{2}$
\cite{bose_quantum_2003}, which is the state-transfer fidelity averaged
over all possible input states $\vert\psi_{0}\rangle$. In the interaction
picture, $f_{0,N+1}(T)=\left|_{S}\left\langle \psi\right|\rho_{S}(T)\left|\psi\right\rangle _{S}\right|$
where $\vert\psi\rangle_{S}=\vert1_{0}\rangle\otimes\vert0_{z}0_{N+1}\rangle{}_{S}$
and $\vert\psi\rangle_{S}\otimes\vert\psi\rangle_{B}$ is the initial
state of $S+B$.

In the ideal regime of an isolated 3-level system, perfect state transfer
occurs when the accumulated phase due to the modulation control 
\begin{equation}
\phi(T)=\tilde{J_{z}}\int_{0}^{T}\alpha(t)dt\label{eq:phi_phase}
\end{equation}
 satisfies $\phi(T)=\frac{\pi}{\sqrt{2}}$. Obviously, this condition
does not strictly hold when the system-bath interaction is accounted
for, yet it is still adequate within the second-order approximation
in $H_{SB}$ used in Eq. (\ref{eq:rho_s}). In this approximation,
$f_{0,N+1}(T)$ takes the form 
\begin{equation}
f_{0,N+1}(T)=1-\zeta(T),\label{eq:f_0,N+1}
\end{equation}
where 
\begin{equation}
\zeta(T)=\Re\int_{0}^{T}\!\! dt\!\int_{0}^{t}\! dt'\underset{\pm}{\sum}\Omega_{\pm}(t)\Omega_{\pm}(t')\Phi_{\pm}(t-t')).\label{eq:eta_t}
\end{equation}
Here, $\Phi_{\pm}(t)=\sum_{k\in k{}_{odd(even)}}|\tilde{J}_{k}|^{2}e^{-i\omega_{k}t}$
are the bath-correlation functions, while $\Omega_{+}(t)=\alpha(t)cos(\sqrt{2}\phi(t))$
and $\Omega_{-}(t)=\alpha(t)$ are the corresponding dynamical-control
functions (\ref{sec:Appendix-A:-Interaction}-\ref{sec:Appendix-B:-Interaction}).
In the calculations we considered $\vert\psi\rangle_{B}=\vert0\rangle_{B}$.
However, in the weak-coupling regime the transfer fidelity remains
the same for a completely unpolarized state \cite{danieli_quantum_2005,yao_robust_2011}
or any other initial state \cite{ping_practicality_2013} of the bath.

In the energy domain, Eq. $\!$(\ref{eq:eta_t}) has the convolutionless
form \cite{escher_optimized_2011,bensky_optimizing_2012,petrosyan_reversible_2009,gordon_universal_2007,gordon_optimal_2008,kofman_universal_2001,kofman_unified_2004}
\begin{equation}
\zeta(T)=\int_{-\infty}^{\infty}d\omega\underset{\pm}{\sum}F_{T,\pm}(\omega)G_{\pm}(\omega),\label{eq:eta_w}
\end{equation}
where the Fourier-transforms 
\begin{equation}
G_{\pm}(\omega)=\frac{1}{2\pi}\int_{-\infty}^{\infty}dt\Phi_{\pm}(t)e^{i\omega t},\: F_{T,\pm}(\omega)=\frac{1}{2\pi}|\int_{0}^{T}dt\Omega_{\pm}(t)e^{i\omega t}|^{2}\label{eq:G(w)  F(w)}
\end{equation}
 are the bath-correlation spectra, $G_{\pm}(\omega)$, associated
with odd(even) parity modes and the spectral filter functions, $F_{T,\pm}(\omega)$,
which can be designed by the modulation control.

\begin{figure}
\centering{}\includegraphics[width=0.4\columnwidth]{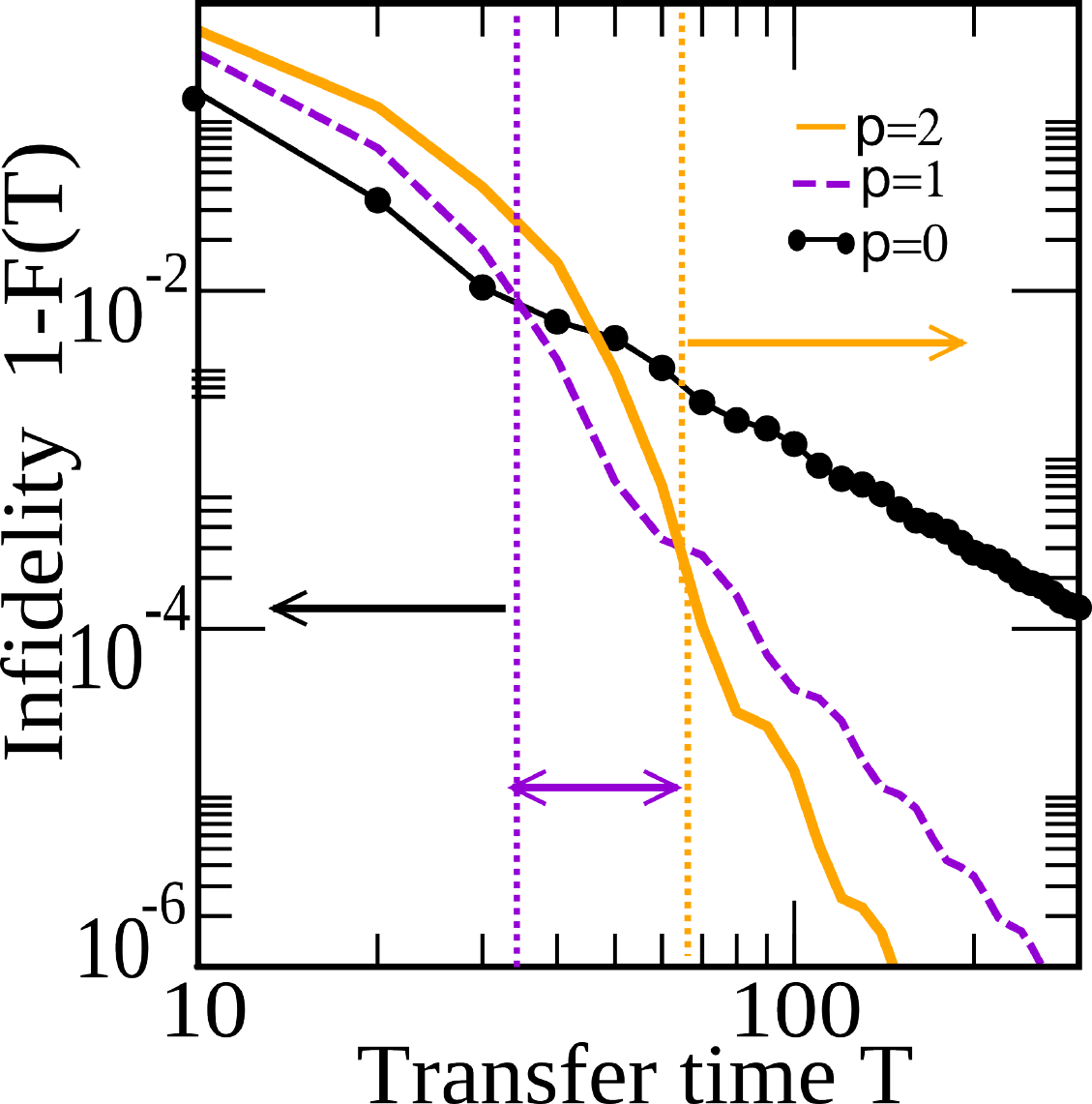}\caption{\label{fig:infidelity-Gw-semicircle}(Color online) Infidelity $1-F(T)$
as a function of transfer time $T$ under optimal control $\alpha_{p}(t)$:
$p=0\mbox{ (black dotted), \ensuremath{p=1}(violet dashed), }p=2$
(orange thin), when the noise or bath-correlation spectrum $G(\omega)$
is bounded by the Wigner-semicircle with a central gap around $\omega_{z}$
(\ref{sec:Appendix-C:-Considerations}).}
\end{figure}

\section{\label{sec:Optimization-method}Optimization method}

To ensure the best possible state-transfer fidelity, we use modulation
as a tool to minimize the infidelity $\zeta(T)$ in (\ref{eq:eta_t}-\ref{eq:eta_w})
by rendering the overlap between the interacting bath- and filter-spectrum
functions as small as possible \cite{clausen_bath-optimized_2010,clausen_task-optimized_2012}.

\subsection{\label{sub:Optimizing-the-modulation}Optimizing the modulation control
for non-Markovian baths}

The minimization of $\zeta(T)$ in (\ref{eq:eta_t}) can be done for
a specific bath-correlation function of a given channel which represents
a non-Markovian bath. The Euler-Lagrange (E-L) equation for minimizing
$\zeta(T)$ with the energy constraint 
\begin{equation}
E(T)=\tilde{J}_{z}^{2}\int_{0}^{T}|\alpha(t)|^{2}dt\label{eq:Energy}
\end{equation}
 turns out to be 
\begin{equation}
\frac{d}{dt}(\frac{\partial\zeta}{\partial\dot{\phi}}-\lambda\frac{\partial E}{\partial\dot{\phi}})-(\frac{\partial\zeta}{\partial\phi}-\lambda\frac{\partial E}{\partial\phi})=0,\label{eq:E-L}
\end{equation}
where $\lambda$ is the Lagrange multiplier and $\dot{\phi}=\tilde{J_{z}}\alpha$.
The optimal modulation can be obtained by solving the integro-differential
equation 
\begin{equation}
\begin{array}{cc}
\ddot{\phi}(t) & =\frac{\sqrt{E}Q(t,\phi(t),\dot{\phi}(t))}{\tilde{J}_{z}\sqrt{\int_{0}^{T}dt\left|\int_{0}^{t}dt'Q(t',\phi(t'),\dot{\phi}(t'))\right|^{2}}},\end{array}\label{eq:phi''(E)}
\end{equation}
where 
\begin{equation}
\begin{array}{l}
Q(t,\phi(t),\dot{\phi}(t))=\int_{0}^{T}dt'\Theta(t-t')\frac{\dot{\phi}(t')}{2\tilde{J}_{z}^{4}}\!\left(\frac{d\Phi_{+}(t-t')}{dt}cos(\sqrt{2}\phi(t))cos(\sqrt{2}\phi(t'))+\frac{d\Phi_{-}(t-t')}{dt}\right)\\
\;\;\;\;\;\;\;\;\;\;\;\;\;\;\;\;\;\;\;\;\;\;\;\;\;\;\;+\frac{\dot{\phi}(t)}{2\tilde{J}_{z}^{4}}\left(\Phi_{+}(0)cos^{2}(\sqrt{2}\phi(t))+\Phi_{-}(0)\right).
\end{array}\label{eq:Q}
\end{equation}
The solution of Eq. (\ref{eq:phi''(E)}) should satisfy the boundary
conditions $\phi(0)=0$ and $\phi(T)=\frac{\pi}{\sqrt{2}}$ to ensure
the required state transfer.

In general the bath-correlations have recurrences and time fluctuations
due to mesoscopic revivals in finite-length channels. Therefore, it
is not trivial to solve Eqs. (\ref{eq:phi''(E)}-\ref{eq:Q}) analytically
rather than solving them numerically for each specific channel. We
however are interested in obtaining universal analytical solutions
for state-transfer in the presence of non-Markovian noise sources.
To this end, we here discuss suitable criteria for optimizing the
state transfer in such cases.

We require the channel to be symmetric with respect to the source
and target qubits and the number of eigenvalues to be odd. These requirements
allow for a central eigenvalue that is \textit{invariant under noise}
on the couplings. This holds provided a \textit{gap exists} between
the central eigenvalue and the adjacent ones, \textit{i.e.} they are
not strongly blurred (mixed) by the noise, so as not to make them
overlap. At the same time, we assume that the discreteness of the
bath spectrum of the quantum channel is smoothed out by the noise,
since it tends to affect more strongly the higher frequencies \cite{zwick_robustness_2011,zwick_spin_2012,Zwick_Chapt_2013}.
Then, if we consider the central eigenvalue as part of the system,
a common characteristic of $G_{\pm}(\omega)$ is to have a central
gap (as exemplified in Fig. \ref{fig:chain-FilterFunction}b). 

Therefore, in order to minimize the overlap between $G_{\pm}(\omega)$
and $F_{T,\pm}(\omega)$ for general gapped baths, and thereby the
transfer infidelity in (\ref{eq:eta_w}), we will design a narrow
bandpass filter centered on the gap.

We present a universal approach that allows us to obtain analytical
solutions for a narrow bandpass filter around $\omega_{z}$. Since
$G_{-}(\omega)$ has a narrower gap than $G_{+}(\omega)$, we optimize
the filter $F_{T,-}(\omega)$ under the variational E-L method. We
seek a narrow bandpass filter, whose form on time-domain via Fourier-transform
decays as slowly as possible, so as to filter out the higher frequencies.
This amounts to maximizing 
\begin{equation}
F_{T,-}(\tau)=\int_{-\infty}^{\infty}F_{T,-}(\omega)e^{-i\omega\tau}d\omega=\int_{0}^{T}\alpha(t)\alpha(t+\tau)dt,\label{eq:Fiter_tau}
\end{equation}
subject to the variational E-L equation (\ref{eq:E-L}), upon replacing
$\zeta$ by $F_{T,-}$. Since there is no explicit dependence on $\phi$,
the second term therein is null, $\frac{\partial}{\partial\phi}(F_{T,-}-\lambda_{E}E)=0$,
yielding 
\begin{equation}
\alpha(t+\tau)+\alpha(t-\tau)=\lambda_{E}\alpha(t)+\lambda_{\phi},\label{eq:alpha_tau}
\end{equation}
where $\lambda_{E}$ is the Lagrange multiplier and $\lambda_{\phi}$
is an integration constant chosen to satisfy the boundary conditions
obeyed by the accumulated phase (\ref{eq:phi_phase}).

Analytical solutions of (\ref{eq:alpha_tau}) are obtainable for small
$\tau$, corresponding to the differential equation 
\begin{equation}
\overset{..}{\alpha}(t)=-\tilde{\lambda}_{E}\alpha(t)+\tilde{\lambda}_{\phi},\label{eq:alpha_dif_Eq}
\end{equation}
 with $\tilde{\lambda}_{E}=\frac{-(\lambda_{E}-2)}{\tau^{2}}$ and
$\tilde{\lambda}_{\phi}=\frac{\lambda_{\phi}}{\tau^{2}}$ . It has
a general solution 
\begin{equation}
\alpha(t)=Asin(\omega_{v}t)+Bcos(\omega_{v}t)+C.\label{eq:alpha_gral_sol}
\end{equation}
The unknown parameters are then optimized under chosen constraints,
e.g. on the boundary coupling, the transfer time, the energy, etc. 

The frequencies $\omega_{v}$ that give a low and flat filter $F_{T,-}(\omega)$
outside a small range around $\omega=\omega_{z}=0$ are $\omega_{v}=\frac{\pi n}{T}$,
$n\epsilon\mathbb{Z}$, since the components of $\alpha(t)$ that
oscillate with $\omega_{v}$ then interfere destructively. Only if
$n=0,1,2$ will the filter have a \textit{single} central peak around
$\omega=0$, and the contribution of larger frequencies will be suppressed,
while the filter-overlap with the central energy level will be maximized;
for larger $n$, the central peak splits and additional peaks appear
at larger frequencies.

Therefore, the analytical expressions for the optimal solutions satisfying
$\phi(0)=0$ and $\phi(T)=\frac{\pi}{\sqrt{2}}$ are found to be
\begin{equation}
\alpha_{p}(t)=\alpha_{M}sin^{p}\left(\frac{\pi t}{T}\right),\label{eq:Opt Mod alphap}
\end{equation}
where $p=0,1,2$\textcolor{black}{,
\begin{equation}
\alpha_{M}=c_{p}\frac{\pi}{\sqrt{2}\tilde{J}_{z}T}\label{eq:alpha_Mp,Tp}
\end{equation}
}

\noindent and $c_{p}=\frac{\sqrt{\pi}\Gamma(\frac{1+p}{2})}{\Gamma(\frac{1+p}{2})}(c_{0}=1,\, c_{1}=\frac{\pi}{2},\, c_{2}=2)$.
Here $p=0$ means static control, while $p=1,2$ stand for dynamical
control. Note that $T$ and $\alpha_{M}=max\{\alpha_{p}(t)\}$ cannot
be independently chosen. If the transfer time is fixed, then the maximum
amplitude depends on $p$, $\alpha_{M}=\alpha_{M_{p}}$, according
to Eq. (\ref{eq:alpha_Mp,Tp}). Similarly, if the maximum amplitude
is kept constant, then the transfer time will depend on $p$, $T=T_{p}$,
by Eq. (\ref{eq:alpha_Mp,Tp}). 

The different solutions in Eq. (\ref{eq:Opt Mod alphap}) are sinc-like
bandpass filter functions around $0$ that become narrower as $T$
increases. For $p=0$, which satisfies the minimal-energy condition
$E_{min}(T_{0})=\frac{\pi^{2}}{2T_{0}}$, the corresponding filter
is the narrowest around $0$, but it has many wiggles on the filter
tails (Fig. \ref{fig:chain-FilterFunction}b) which overlap with bath-energies
that hamper the transfer. In contrast, the $p=1,2$ bandpass filters
are wider (for the same $T$) and require more energy, $E_{1}=\frac{\pi^{2}}{8}E_{{\scriptstyle min}}$
and $E_{2}=\frac{3}{2}E_{min}$ respectively, but these filters are
flatter and lower throughout the bath-energy domain. 

Hence, the bandpass filter width (\textit{i.e.} full width at half
maximum) and the overlap of its tail-wiggles with bath-energies as
a function of $T$, determine which modulations $\alpha_{p}(t)$ are
optimal, as shown in the inset of Fig. \ref{fig:chain-FilterFunction}b
($F_{T,+}(\omega)$ filters out a similar spectral range). The shorter
$T$, the lower is $p$ that yields the highest fidelity, because
the central peak of the filter that produces the dominant overlap
with the bath spectrum is then the narrowest. However, as $T$ increases,
larger $p$ will give rise to higher fidelity, because now the tails
of the filter make the dominant contribution to the overlap. As shown
in Fig. \ref{fig:infidelity-Gw-semicircle}, the filter for $p=1,2$
can improve the transfer fidelity \textit{by orders of magnitude}
in a noisy gapped bath bounded by the Wigner-semicircle, which is
representative of fully randomized channels \cite{wigner_distribution_1958}
(\ref{sec:Appendix-C:-Considerations}).

\subsection{Optimizing the modulation control for a Markovian Bath}

We next consider the worst-case scenario of a Markovian bath, where
the bath-correlation functions $\Phi_{\pm}(\tau)$ vanish for $\tau>0$.
This is the case when the gap is closed by a noise causing the bath
energy levels to fluctuate faster than the system dynamics. We note
that, finding optimal solutions for the noise spectrum of a Markovian
bath is important for the case where the gap is reduced or even lost
in static cases. 

The infidelity function (\ref{eq:eta_t}) that must be minimized when
the correlation time $\tau_{c}=0$, \textit{i.e.} $\Phi_{\pm}(\tau)=\delta(\tau)$,
is 
\begin{equation}
\zeta(T)=\Re\int_{0}^{T}dt\frac{\dot{\phi}^{2}(t)}{\tilde{J}_{z}^{2}}\bigl(\Phi_{+}(0)cos^{2}(\sqrt{2}\phi(t))+\Phi_{-}(0)\bigr).\label{eq:Mark}
\end{equation}
The E-L equation under energy constraint (\ref{eq:E-L}), is now 
\begin{equation}
\ddot{\phi}(t)\!\left(\Phi_{+}(0)cos^{2}(\sqrt{2}\phi(t))\!+\!\Phi_{-}(0)\!-\!2\lambda\tilde{J}_{z}^{2}\right)\!-\!\sqrt{2}\dot{\phi}^{2}(t)\,\Phi_{+}(0)cos(\sqrt{2}\phi(t))sin(\sqrt{2}\phi(t))\!=\!0.\label{eq:phi Mark bath-1-1}
\end{equation}
This equation has a non-trivial analytical solution and the modulation
that minimizes $\zeta(T)$ is given by the following transcendental
equation 
\begin{equation}
{\normalcolor \begin{array}{c}
T\intop_{0}^{\phi(t)}\sqrt{cos(2\sqrt{2}\phi)\Phi_{+}(0)+\Phi_{+}(0)+2\Phi_{-}(0)-2\lambda\tilde{J}_{z}^{2}}d\phi\\
-t\intop_{0}^{\phi(T)}\sqrt{2(\Phi_{+}(0)cos^{2}(\sqrt{2}\phi)+\Phi_{-}(0)-\lambda\tilde{J}_{z}^{2})}d\phi=0.
\end{array}}\label{eq:analitycal solution Mark Bath}
\end{equation}
The infidelity for this optimal modulation almost coincides with the
one obtained for static control ($\alpha_{p=0}(t)=\alpha_{M}$ from
Eq. (\ref{eq:Opt Mod alphap})), \textit{i.e.} 
\begin{equation}
1-F(T)\approx\frac{\pi^{2}N}{6\sqrt{2}JT}(1-\frac{\pi^{2}N}{16\sqrt{2}JT}),\: T=\frac{\pi\sqrt{N}}{2\alpha_{M}J},
\end{equation}
 and they only differ by about 0.1\%. This optimal modulation can
be phenomenologically approximated by 
\begin{equation}
\alpha(t)\approx a\alpha_{M}+b\, sin^{q}(\frac{t\pi}{T})),\: q\sim3.5,\:\frac{b}{a}\sim\frac{1}{3},\: a\sim0.84,
\end{equation}
assuming no constraints (\textit{$\lambda=0$}). An example of the
performance of this solution is discussed below and shown in Fig.
\ref{fig:Mark-Noise}.

\begin{figure}
\centering{}\includegraphics[width=0.7\columnwidth]{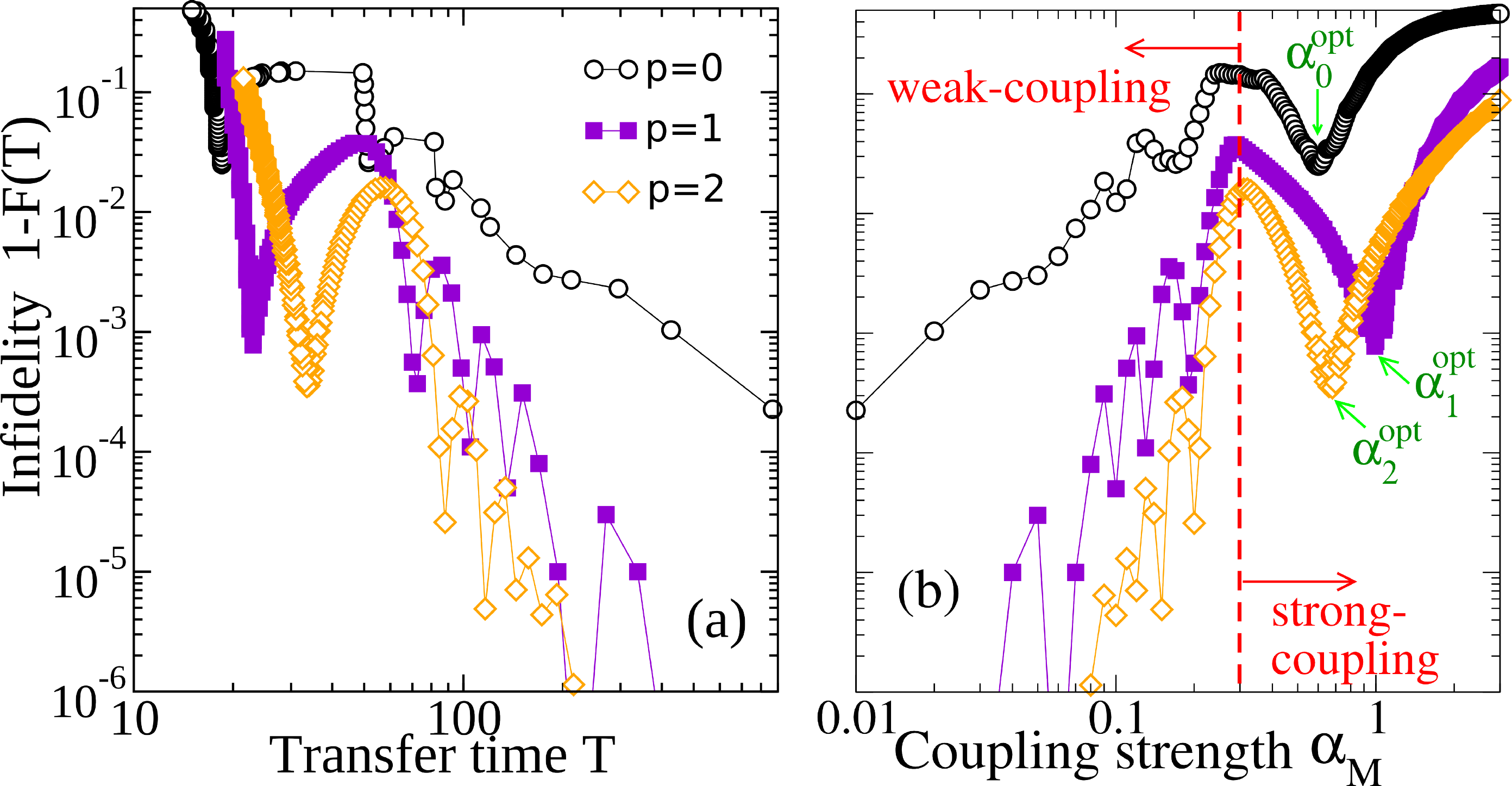}\caption{\label{fig:Fmax-alp}(Color online) Transfer infidelity $1-F(T)$
for a modulated boundary-controlled coupling $\alpha_{p}(t)=\alpha_{M}sin^{p}(\frac{\pi t}{T})$
as a function of (a) the transfer time $T$, (b) the maximum value
of the boundary coupling $\alpha_{M}$: $p=0$ (empty circles), $p=1$
(filled squares), $p=2$ control (empty diamonds). The quantum channel
is a homogeneous spin-chain with $N+2=31$ spins and $J=1$.}
\end{figure}

\section{\label{sec:Optimal-control-of}Optimal control of transfer in a homogeneous
spin-chain channel }

Consider a \textit{uniform} (homogeneous) spin-chain channel, \textit{i.e.}
$J_{i}\equiv J$ in Eq. (\ref{eq:hamiltonian}), whose energy eigenvalues
are $\omega_{k}=2Jcos(\frac{k\pi}{N+1})$ \cite{wojcik_unmodulated_2005}.
In Fig. \ref{fig:Fmax-alp}, we show the performance of the general
optimal solutions (\ref{eq:Opt Mod alphap}) for this specific channel
as a function of $\alpha_{M}$ and $T$. 

The approach based on Eq. (\ref{eq:eta_t}) strictly holds in the
weak-coupling regime $(\alpha_{M}\ll1)$ \cite{clausen_bath-optimized_2010,clausen_task-optimized_2012,escher_optimized_2011,bensky_optimizing_2012,petrosyan_reversible_2009,kofman_universal_2001,kofman_unified_2004}.
In this regime (marked with arrows in Fig. \ref{fig:Fmax-alp}b),
we found that the transfer time is $T_{p}\!\approx\! c_{p}\frac{\pi\sqrt{N}}{2\alpha_{M}J}$,
and the infidelity decreases by reducing $\alpha_{M}$ according to
a power law, aside from the oscillations due to the discrete nature
of the bath-spectrum (see \ref{sec:Appendix-C:-Considerations}).
The filter tails are sinc-like functions, so that when a zero of the
filter matches a bath-energy eigenvalue, the infidelity exhibits a
dip. Aside from oscillations, the best tradeoff between speed and
fidelity within this regime is given by the optimal modulation with
$p=2$ (for the system described in Fig. \ref{fig:Fmax-alp}a). 

However, this approach can also be extended \textit{to strong couplings}
$\alpha_{M}$, since it \textit{becomes compatible with the weak-coupling
regime} under the optimal filtering process that increases the state
fidelity in the interaction picture \cite{gordon_universal_2007,gordon_dynamical_2009,kurizki_universal_2013}.
The bandpass filter width increases as $T$ decreases; consequently,
in the strong coupling regime $(\alpha_{M}\sim1)$ the filter may
now overlap the bath energies closest to $\omega_{z}$, but still
block the higher bath energies, which are the most detrimental for
the state transfer \cite{zwick_robustness_2011,zwick_spin_2012,Zwick_Chapt_2013}.
Then, the participation of the closest bath energies yields a transfer
time $T_{p}\!\approx\! c_{p}\frac{N}{2J}$. There is a clear minimal
infidelity value at the point that we denote as $\alpha_{M_{p}}^{opt}$
which depends on $p$ (Fig. \ref{fig:Fmax-alp}b); thus extending
the previous static-control ($p=0$) results, where an optimal $\alpha_{M_{0}}^{opt}$
was found \cite{zwick_quantum_2011,banchi_long_2011,banchi_optimal_2010,zwick_spin_2012}.
The infidelity dip corresponds to a better filtering-out (suppression)
of the higher energies, retaining only those that correspond to an
almost equidistant spectrum of $\omega_{k}$ around $\omega_{z}$,
which allow for coherent transfer \cite{zwick_spin_2012}.

Figure \ref{fig:Fmax-alp}b shows that by fixing $\alpha_{M}$, the
dynamical control ($p=1,2$) of the boundary-couplings reduces the
transfer infidelity \textit{by orders of magnitude} only at the expense
of slowing down the transfer time $T_{p}$ at most by a factor of
2, $\frac{T_{p}}{T_{0}}\approx\frac{c_{p}}{c_{0}}\leq2$. If the constraint
on $\alpha_{M}$ can be relaxed, \textit{i.e.} more energy can be
used, the advantages of dynamical control can be even more appreciated
for both infidelity decrease and transfer-time reduction by orders
of magnitude, as shown in Fig. \ref{fig:Fmax-alp}a. Hence, our main
result is that the speed-fidelity tradeoff can be drastically improved
under optimal dynamical control.

\section{Robustness against different noises}

We now explicitly consider the effects of optimal control on noise
affecting the coupling strengths, also called off-diagonal noise,
causing: $J_{i}\rightarrow J_{i}+J_{i}\Delta_{i}(t),\; i=1,...,N$
with $\Delta_{i}$ being a uniformly distributed random variable in
the interval $\left[-\varepsilon_{J},\varepsilon_{J}\right]$. Here
$\varepsilon_{J}>0$ characterizes the noise or disorder strength.
When $\Delta_{i}$ is time-independent, it is called \textit{static
noise}, as was considered in other state-transfer protocols \cite{de_chiara_perfect_2005,ronke_effect_2011,zwick_robustness_2011,zwick_spin_2012}.
When $\Delta_{i}(t)$ is time-dependent, we call it \textit{fluctuating
noise} \cite{Burgarth_fluctuating}\textit{.} These kinds of noises
will affect the bath energy levels, while the central energy $\omega_{z}$
remains invariant \cite{zwick_robustness_2011,Zwick_Chapt_2013}.
In the following we analyse the performance of the control solutions
obtained in Sec. \ref{sec:Optimization-method} for these types of
noise and later on, in Sec. \ref{sub:Other-sources-of} we discuss
briefly the effects of other sources of noise.

\begin{figure}
\centering{}\includegraphics[width=0.36\columnwidth]{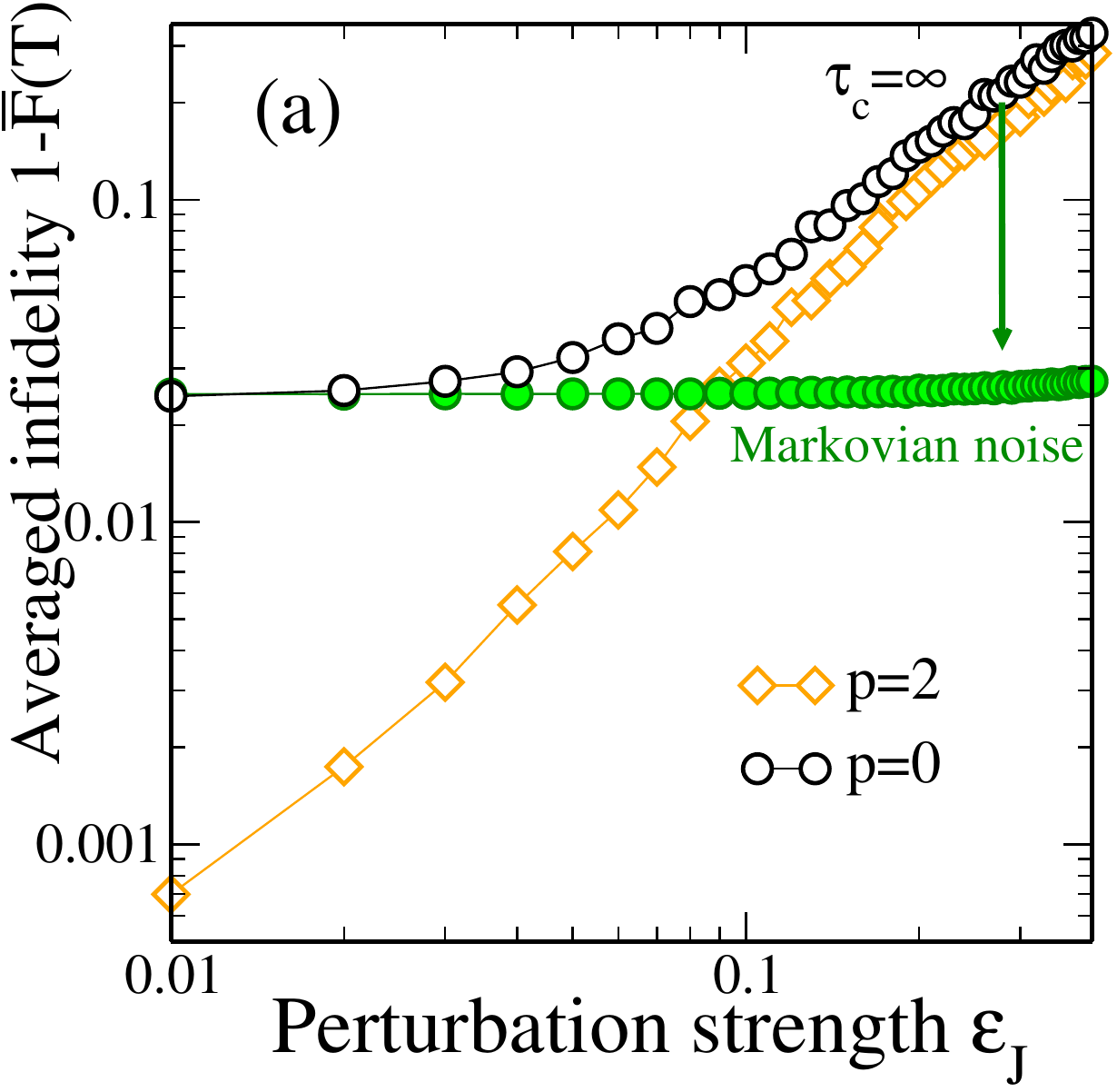}\includegraphics[width=0.35\columnwidth]{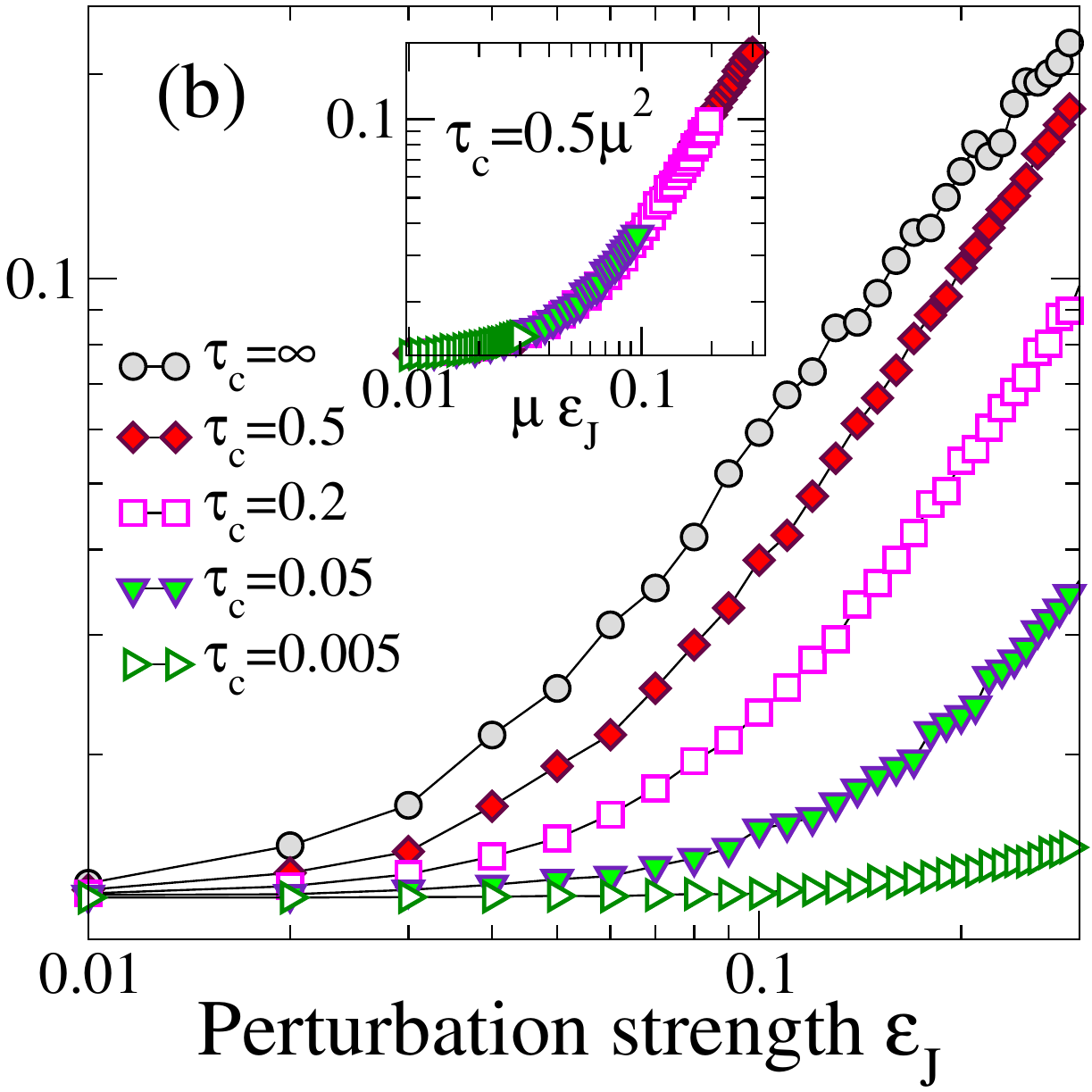}\caption{\label{fig:Noise}(Color online) Transfer infidelity for a modulated
boundary-controlled coupling $\alpha_{p}(t)$ as a function of the
perturbation strength $\varepsilon_{J}$ of the noisy homogeneous
channel, averaged over \foreignlanguage{american}{$N_{av}$} noise
realizations for (a) static and fluctuating noisy channels $\alpha_{M_{0}}^{opt}=0.6$,
$\alpha_{M_{2}}^{opt}=0.7$ and $N_{av}=10^{3}$. In static noisy
channels, the infidelity obtained under static control $p=0$ (empty
circles) is shown to be strongly reduced when dynamical $p=2$ control
is applied (empty squares). A fluctuating noisy channel is less damaging;
in the Markovian limit, where the correlation time of the noise fluctuations
$\tau_{c}\rightarrow0$ ($p=0$, green solid circles), the infidelity
converges to its unperturbed value. The homogeneous channel has $N+2=31$
spins and $J=1$. (b) Same plot for fluctuating noise, ranging between
static and Markovian noise for $\alpha_{M_{0}}=0.1$, $T=88$ and
$N_{av}=200$. Here $\tau_{c}$ is the correlation time of the noise
fluctuations (see text). Faster fluctuations reduce the noise effect
and thereby the fidelity decay. The inset shows the scaling of the
infidelity, where the effective noise strength is scaled with $\mu=\sqrt{2\tau_{c}}$
(all curves overlap). The scaled noise strength depends on the noise
correlation time $\tau_{c}$.}
\end{figure}

\subsection{\textit{Static noise} }

Static control on the boundary-couplings can suppress static noise
\cite{zwick_robustness_2011,zwick_spin_2012} but here we show that
dynamical boundary-control makes the channel even more robust, because
it filters out the bath-energies that damage the transfer. To illustrate
this point, we compare the effect of modulations $\alpha_{p}(t)$
with $\alpha_{M}=\alpha_{M_{p}}^{opt}$ for $p=0$ and 2 in the strong-coupling
regime (Fig. \ref{fig:Noise}a). There is an evident advantage of
dynamical control with $p=2$ compared to static control ($p=0$),
at the expense of increasing the transfer time by only a factor of
2, $\frac{T_{2}}{T_{0}}\approx2$. In the weak-coupling regime, if
we choose $\alpha_{M}$ such that the transfer fidelity is similar
for $p=0$ and $p=2$, then both cases are similarly robust under
static disorder, but the modulated case $p=2$ is an order of magnitude
faster. Remarkably, because of disorder-induced localization \cite{Porter1965,Imry2002,akulin_spectral_1993,pellegrin_mie_2001},
regardless of how small is $\alpha_{M}$, the averaged fidelity under
static noise cannot be improved beyond the bound 
\begin{equation}
1-\bar{F}\propto N\varepsilon_{J}^{2},\:(\varepsilon_{J}\ll1).
\end{equation}

\subsection{\textit{Markovian noise}}

\begin{figure}
\centering{}\includegraphics[width=0.4\columnwidth]{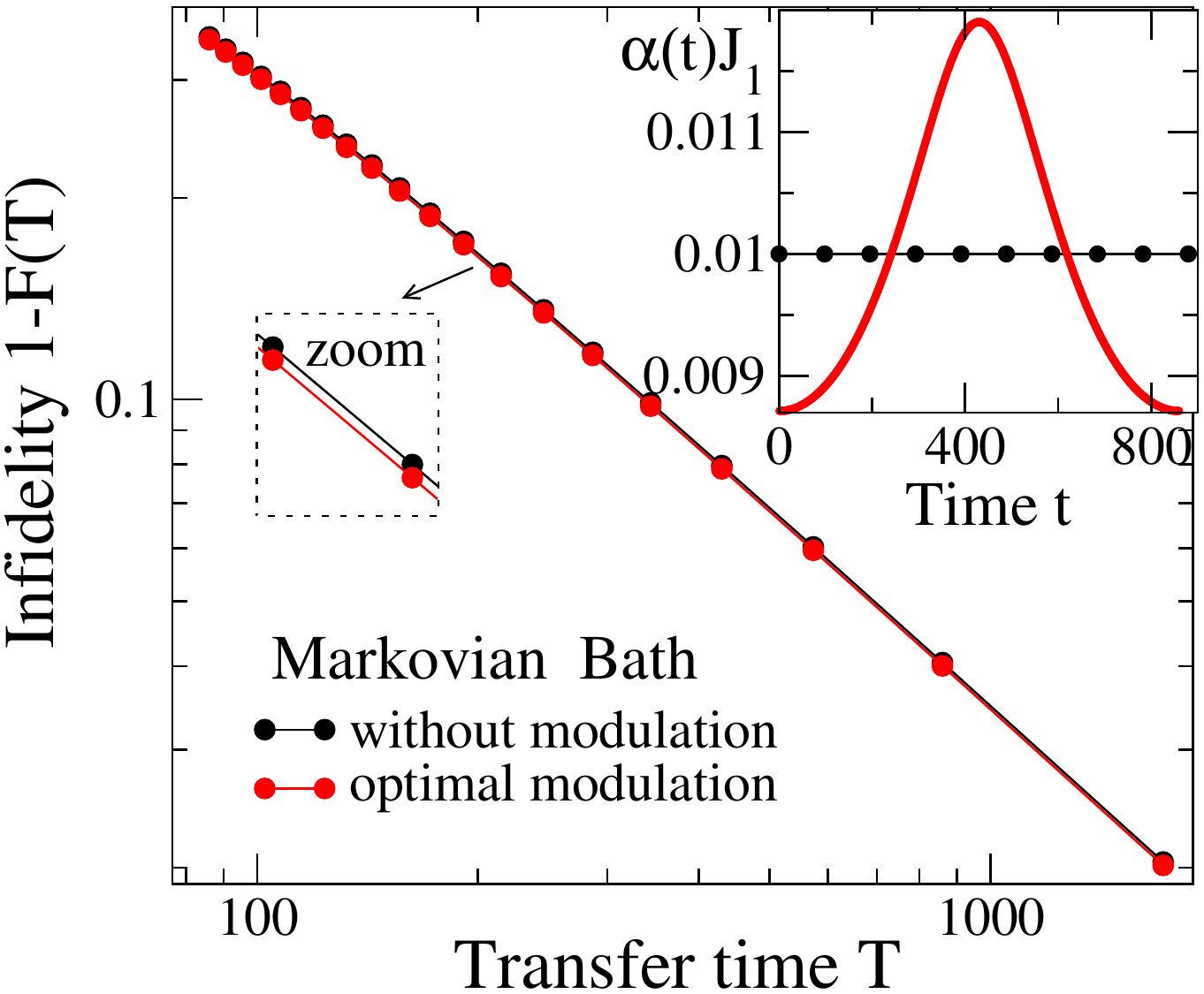}\caption{\label{fig:Mark-Noise}(Color online) Transfer infidelity as a function
of the transfer time $T$ for the optimal control solution in the
case of a Markovian bath (Eq. (\ref{eq:analitycal solution Mark Bath}))
and without modulation ($p=0$). $N+2=31$, $J_{1}=J_{N}$. The corresponding
boundary-couplings for $T=860$ are shown in the inset. }
\end{figure}

The worst scenario for quantum state transfer is the absence of an
energy gap around $\omega_{z}$. This case corresponds to Markovian
noise characterized by $\left\langle \Delta_{i}(t)\Delta_{i}(t+\tau)\right\rangle =\delta(\tau)$,
where the brackets denote the noise ensemble average, or equivalent
to a bath correlation-function that vanishes at $\tau>0$. In this
case there is an analytical solution for the optimal modulation given
by Eq. (\ref{eq:analitycal solution Mark Bath}), although the infidelity
achieved by it almost coincides with the one obtained by the static
($p=0$) optimal control (Fig. \ref{fig:Mark-Noise}). Counterintuitively,
arbitrarily high fidelities can be achieved for such noise by decreasing
$max\left|\alpha(t)\right|$ and thereby slowing down the transfer.
This comes about because in a Markovian bath, the very fast coupling
fluctuations suppress the disorder-localization effects that hamper
the transfer fidelity as we show below for a typical case.

\subsection{\textit{Non-Markovian noise}}

We now consider a non-Markovian noise of the form $J_{i}+J_{i}\Delta_{i}(t)$,
where $\Delta_{i}(t)=\Delta_{i}\left(\left[t/\tau_{c}\right]\right)$,
where the integer part $\left[t/\tau_{c}\right]=n$ defines a noise
$\Delta_{i}\left(n\right)$ that randomly varies between the interval
$\left[-\varepsilon_{J},\varepsilon_{J}\right]$ at time-intervals
of $\tau_{c}$ during the transfer. We observe a convergence of the
transfer fidelity to its value without noise as the noise correlation
time $\tau_{c}$ decreases (Fig. \ref{fig:Noise}b). Consequently
the fidelity can be substantially improved by reducing $\alpha_{M}$.
The effective noise strength scales down as $\tau_{c}^{1/2}$ (Fig.
\ref{fig:Noise}b, inset). By contrast to the Markovian limit $\tau_{c}\rightarrow0$,
dynamical control can strongly reduce the infidelity in the non-Markovian
regime that lies between the static and Markovian limits and whose
bath-spectrum is gapped.

\subsection{\textit{\label{sub:Other-sources-of}Other sources of noise}}

\textit{Timing errors}: In addition to resilience to noise affecting
the spin-spin couplings, there is another important characteristic
of the transfer robustness, namely, the length of the time window
in which high fidelity is obtained. The fidelity $F(t)$ under optimal
dynamical control ($p=1,2)$, yields a wider time-window around $T$
where the fidelity remains high compared with its static ($p=0)$
counterpart. This allows more time for determining the transferred
state or using it for further processing. Consequently, the \textit{robustness
}against timing imperfections \cite{kay_perfect_2006,zwick_robustness_2011}
\textit{is increased} under optimal dynamical control.

\noindent \textit{On-site energy noise}: This kind of noise, alias
diagonal-noise, can be either static or fluctuating. The static one
can give rise to the emergence of quasi-degenerate central states.
Then, the dynamical control approach introduced in this work is still
capable of isolating the ``system'' defined here (Sec. \ref{sec:Quantum channel and state transfer fidelity})
from the remaining ``bath'' levels. It may happen that the spin
network is not symmetric with respect to the source and target spins,
and then the effective couplings of the source and target qubits with
the central level will not be symmetric. This asymmetry can be effectively
eliminated by boundary control. On the other hand, a fluctuating diagonal-noise
that may produce a fluctuation of the central energy level is here
fought by optimizing the tradeoff between speed and fidelity as detailed
above. Additional dynamical control of only the source and target
spins can be applied to avoid these decoherence effects, by the mapping
to an effective 3-level system, as a variant of dynamical decoupling
\cite{Viola_Dynamical_1998,viola_dynamical_1999,Viola_RobustDD_2003,Lidar_QDynDec_2005}.

\section{Conclusions}

We have proposed a general, optimal dynamical control of the tradeoff
between the speed and fidelity of qubit-state transfer through the
central-energy global mode of a quantum channel in the presence of
either static or fluctuating noise. Dynamical boundary-control has
been used to design an optimal spectral filter realizable by universal,
simple, modulation shapes. The resulting transfer infidelity and/or
transfer time can be reduced by orders of magnitude, while their robustness
against noise on the spin-spin couplings is maintained or even improved.
Transfer-speed maximization is particularly important in our strive
to reduce the random phase accumulated during the transfer when energy
fluctuations (diagonal noise) affect the spins \cite{Ajoy_perfect_2013}.
We have shown that, counterintuitively, static noise is more detrimental
than fluctuating noise on the spin-spin couplings. This general approach
is applicable to quantum channels that can be mapped to Hamiltonians
quadratic in bosonic or fermionic operators \cite{cappellaro_dynamics_2007,rufeil-fiori_effective_2009,doronin_multiple-quantum_2000,yao_robust_2011,yao_quantum_2013}.
We note that our control is complementary to the recently suggested
control aimed at balancing possible asymmetric detunings of the boundary
qubits from the channel resonance \cite{Ajoy_perfect_2013,yao_quantum_2013}.

\ack We acknowledge the support of ISF-FIRST (Bikura) and the EC
Marie Curie (Intra-European) Fellowship (G.A.A.).

\appendix

\section{The Hamiltonian in the interaction picture\label{sec:Appendix-A:-Interaction}}

The system-bath Hamiltonian (Eq. (\ref{eq:Hsb}) of the main text)
splits into a sum of symmetric and antisymmetric system operators
that are coupled to odd- and even-bath modes: $H_{SB}(t)=\underset{j=1}{\overset{4}{\sum}}\tilde{S}_{j}\otimes\tilde{B}_{j}^{\dagger},$
where $\tilde{S}_{1(3)}=\alpha(t)(c_{0}+(-)c_{N+1})$, $\tilde{S}_{2(4)}=\tilde{S}_{1(3)}^{\dagger}$,
$\tilde{B}_{1(3)}=\underset{k\in k_{odd(even)}}{\sum}\tilde{J}_{k}b_{k}$
and $\tilde{B}_{2(4)}=\tilde{B}_{1(3)}^{\dagger}$. In the interaction
picture $H_{SB}(t)$ becomes 
\begin{equation}
H_{SB}^{I}(t)=\sum_{j=1}^{4}S_{j}(t)\otimes B_{j}^{\dagger}(t),\label{eq:HI_SB}
\end{equation}
where 
\begin{equation}
\begin{array}{c}
S_{j}(t)=U_{S}^{\dagger}(t)\tilde{S}_{j}(t)U_{S}(t),\, U_{S}(t)=\mathcal{T}e^{-i\intop_{0}^{t}dt^{'}H_{S}(t^{'})},\\
B_{j}(t)=U_{B}^{\dagger}(t)\tilde{B}_{j}U_{B}(t),\, U_{B}(t)=e^{-iH_{B}t};
\end{array}
\end{equation}
and the evolution operators are 
\begin{equation}
\begin{array}{l}
\begin{array}{cl}
U_{S}(t)= & \vert0\rangle_{SS}\langle0\vert+\left(\frac{\cos(\sqrt{2}\phi(t))+1}{2}\right)\left(\vert0\rangle\langle0\vert+\vert N+1\rangle\langle N+1\vert\right)\\
 & +\left(\frac{\cos(\sqrt{2}\phi(t))-1}{2}\right)\left(\vert0\rangle\langle N+1\vert+\vert N+1\rangle\langle0\vert\right)\\
 & +\cos(\sqrt{2}\phi(t))\vert z\rangle\langle z\vert-i\,\frac{\sin(\sqrt{2}\phi(t))}{2}\left(\vert0\rangle\langle z\vert+\vert N+1\rangle\langle z\vert+h.c.\right),\\
U_{B}(t)= & \overset{N}{\underset{k=1,k\neq z}{\sum}}e^{-i\omega_{k}t}\vert k\rangle\langle k\vert+\vert0\rangle_{BB}\langle0\vert,
\end{array}\end{array}
\end{equation}
where the states $\vert0\rangle_{S}=\vert0_{0}0_{z}0_{N+1}\rangle_{S}$
and $\vert0\rangle_{B}=\vert0_{1}...0_{N}\rangle_{B}$ refer to the
zero-excitation states in the system (S) and bath (B) respectively.
Therefore, the bath operators are $B_{1(3)}(t)=\underset{k\in k_{odd(even)}}{\sum}\vert\tilde{J}_{k}\vert^{2}e^{-i\omega_{k}t}\vert k\rangle{}_{B}\langle0\vert,\, B_{2(4)}(t)=B_{1(3)}^{\dagger}(t)$. 

We define a basis of operators $\hat{\nu}_{i}$ to describe the rotating
system operators $S_{j}(t)$ via a rotation-matrix $\Omega_{j,i}(t)$.
They are given by 
\begin{equation}
\begin{array}{cc}
\hat{\nu}_{1}=\vert0\rangle_{S}\left(\langle0\vert+\langle N+1\vert\right) & \hat{\nu}_{2}=\hat{\nu}_{1}^{\dagger},\\
\hat{\nu}_{3}=\vert0\rangle_{S}\langle z\vert & \hat{\nu}_{4}=\hat{\nu}_{3}^{\dagger},\\
\hat{\nu}_{5}=\vert0\rangle_{S}\left(\langle0\vert-\langle N+1\vert\right) & \hat{\nu}_{6}=\hat{\nu}_{5}^{\dagger},
\end{array}\label{eq:nu_i}
\end{equation}
such that $S_{j}(t)=\overset{6}{\underset{i=1}{\sum}}\Omega_{j,i}(t)\hat{\nu}_{i}.$
Given that $S_{1}(t)=\dot{\phi}(t)\left(cos(\sqrt{2}\phi(t))\hat{\nu}_{1}-i\,\sqrt{2}sin(\sqrt{2}\phi(t))\hat{\nu}_{3}\right)$,
$S_{3}(t)=\dot{\phi}(t)\hat{\nu}_{5},\, S_{2(4)}(t)=S_{1(3)}^{\dagger}(t)$
the rotation-matrix vectors are 
\begin{equation}
\begin{array}{c}
\begin{array}{l}
\Omega_{1,i}(t)=\dot{\phi}(t)\left(cos(\sqrt{2}\phi(t)),0,-i\,\sqrt{2}sin(\sqrt{2}\phi(t)),0,0,0\right)\\
\Omega_{2,i}(t)=\dot{\phi}(t)\left(0,cos(\sqrt{2}\phi(t)),0,i\,\sqrt{2}sin(\sqrt{2}\phi(t)),0,0\right)\\
\Omega_{3,i}(t)=\dot{\phi}(t)(0,0,0,0,1,0)\\
\Omega_{4,i}(t)=\dot{\phi}(t)(0,0,0,0,0,1).
\end{array}\end{array}\label{eq:RotationMatrix_Omega}
\end{equation}

\section{The fidelity in the interaction picture\label{sec:Appendix-B:-Interaction}}

Here we derive Eqs. (\ref{eq:f_0,N+1}-\ref{eq:eta_t}) from Eq. (\ref{eq:rho_s})
of the main text. Considering $\vert\psi\rangle=\vert100...0\rangle_{SB}=\vert\psi\rangle_{S}\otimes\vert0\rangle_{B}$
with $\vert\psi\rangle_{S}=\vert1_{0}0_{z}0_{N+1}\rangle_{S}$ as
the initial state, the fidelity is reduced to 
\begin{equation}
\begin{array}{cc}
f_{0,N+1}(T)= & \left|_{S}\left\langle \psi\right|\rho_{S}(T)\left|\psi\right\rangle _{S}\right|=1-\zeta(T)\end{array},
\end{equation}

where $\zeta(T)=T\underset{i,i'=1}{\overset{6}{\sum}}R_{i,i'}(T)\Gamma_{i,i^{'}}$,
with
\begin{equation}
\Gamma_{i,i^{'}}={}_{S}\left\langle \psi\right|[\hat{\nu}_{i},\hat{\nu}_{i'}\left|\psi\right\rangle _{S}{}_{S}\left\langle \psi\right|]\left|\psi\right\rangle _{S}=\delta_{i,2}\delta_{1,i'}+\delta_{i,2}\delta_{5,i'}+\delta_{i,6}\delta_{1,i'}+\delta_{i,6}\delta_{5,i'}
\end{equation}
and 
\begin{equation}
R_{i,i'}(T)\!=\!\frac{1}{T}\int_{0}^{T}dt\!\int_{0}^{t}dt'(\Phi_{2,1}(t-t')\Omega_{2,i}(t)\Omega_{1,i'}(t')+\Phi_{4,3}(t-t')\Omega_{4,i}(t)\Omega_{3,i'}(t')).
\end{equation}
Here $\hat{\nu}_{i'}$ and $\Omega_{j,i}$ are as defined in Eqs.
(\ref{eq:nu_i}-\ref{eq:RotationMatrix_Omega}), while the correlation
functions are 
\begin{equation}
\Phi_{j,j'}(t-t')=\sum_{k\in k_{odd}}\vert\tilde{J}_{k}\vert^{2}e^{-i\omega_{k}(t-t')}\delta_{j,2}\delta_{1,j'}+\sum_{k\in k_{even}}\vert\tilde{J}_{k}\vert^{2}e^{-i\omega_{k}(t-t')}\delta_{j,4}\delta_{3,j'}.
\end{equation}
This leads to the infidelity $\zeta(T)$ of Eq. (\ref{eq:eta_t}).

\section{Considerations for a specific non-Markovian bath: the uniform spin-channel\label{sec:Appendix-C:-Considerations}}

Consider a \textit{uniform} (homogeneous) spin-chain channel, \textit{i.e.}
$J_{i}\equiv J$ in Eq.(\ref{eq:hamiltonian}), whose energy eigenvalues
are $\omega_{k}=2Jcos(\frac{k\pi}{N+1})$. In the weak-coupling regime
where $\alpha_{M}\ll1$, the coupling strength in the interaction
$H_{bc}$, $\tilde{J}_{z}=\sqrt{\frac{2}{N+1}}J$ and $\tilde{J}_{k}=\tilde{J}_{z}sin(\frac{k\pi}{N+1})$,
are always much smaller than the nearest eigenvalue gap $\vert\omega_{z}-\omega_{z\pm1}\vert\sim\frac{2J}{N}$
\cite{wojcik_unmodulated_2005,wojcik_multiuser_2007,yao_robust_2011}.
The correlation function of the bath is 
\begin{equation}
\Phi_{\pm}(\tau)=\underset{k{}_{odd(even)}}{\sum}\left|\sqrt{\frac{2}{N\!+\!1}}Jsin(\frac{k\pi}{N\!+\!1})\right|^{2}e^{-i2Jcos(\frac{k\pi}{N+1})\tau}
\end{equation}
 and has recurrences and time fluctuations due to mesoscopic revivals,
while at short times $t$, it behaves as a Bessel function {\small ${\color{black}{\color{red}{\color{black}\Phi(t)=\frac{2(\alpha_{0}J)^{2}}{J\tau}\mathtt{\mathcal{J}_{1}(}2Jt)}}}$.
}The latter correlation function represents the limiting case of an
infinite channel and it gives a continuous bath-spectrum that becomes
a semicircle. In the case of a finite channel, $G(\omega)$ will be
discrete but modulated by the semicircle with a central gap. If disorder
is considered, the position of the spectrum lines fluctuates from
channel to channel but they are essentially modulated by the semicircle
with a central gap as was considered in the Fig. \ref{fig:chain-FilterFunction}b
of the main text, where 
\begin{equation}
G_{\pm}(\omega)=\frac{1}{2}\sqrt{4J^{2}-\omega^{2}}(1-\Theta(\omega-\omega_{l})\Theta(\omega+\omega_{l})),\:\omega_{l}=\frac{3\omega_{z+1}}{4}.
\end{equation}
This is the Wigner-distribution for fully randomized channels \cite{wigner_distribution_1958}
with a central gap.

\section*{References}

\bibliographystyle{unsrt}
\bibliography{Zwick-ODCST}

\end{document}